\documentclass{aa}
\usepackage{times,float}
\usepackage{epsfig}
\usepackage{graphics}
\usepackage{amssymb}

\begin{document}

\title{The  blue  host  galaxy  of  the  red  GRB~000418  \thanks{Based  on
   observations made  with the Nordic  Optical Telescope, operated  on the
   island of  La Palma  jointly by Denmark,  Finland, Norway, and  Sweden.
   Based on  observations collected at the  European Southern Observatory,
   in La Silla and Paranal (Chile), ESO Large Programmes 165.H-0464(I) and
   265.D-5742(C),  granted  to  the  GRACE  Team.  Based  on  public  data
   collected  under ESO  programme  67.B-0611(A), and  retrieved from ESO
   data archive. Based on HST data collected under programme 8189.}}

\titlerunning{The host galaxy of GRB~000418}
{\small
\author{ J. Gorosabel \inst{1,2,3} \and
  S. Klose \inst{4} \and
  L. Christensen \inst{5} \and
  J.P.U. Fynbo \inst{6,7} \and
  J. Hjorth \inst{7} \and
  J. Greiner \inst{8} \and
  N. Tanvir \inst{9} \and
  B.L. Jensen \inst{7} \and
  H. Pedersen \inst{7} \and
  S.T. Holland \inst{10} \and
  N. Lund \inst{2} \and
  A.O. Jaunsen \inst{11} \and
  J.M. Castro Cer\'on \inst{12} \and
  A.J. Castro-Tirado \inst{1} \and
  A. Fruchter \inst{3} \and
  E. Pian \inst{13} \and
  P. M. Vreeswijk \inst{11} \and
  I. Burud \inst{3} \and
  F. Frontera \inst{14,15} \and
  L. Kaper \inst{16} \and
  C. Kouveliotou \inst{17} \and
  N. Masetti \inst{15} \and
  E. Palazzi \inst{15} \and
  J. Rhoads \inst{3} \and
  E. Rol \inst{16} \and
  I. Salamanca \inst{16} \and
  R.A.M.J. Wijers \inst{16} \and
  E. van den Heuvel \inst{16} } }

\institute{ Instituto de Astrof\'{\i}sica de Andaluc\'{\i}a (IAA-CSIC),
 P.O. Box 03004, E--18080 Granada, Spain; {\tt jgu@iaa.es, ajct@iaa.es}
 \and
  Danish Space Research Institute, Juliane Maries Vej 30, DK--2100
 Copenhagen \O, Denmark; {\tt jgu@dsri.dk, nl@dsri.dk}
 \and
 Space  Telescope Science Institute, 3700  San Martin  Drive, Baltimore, MD
 21218,   USA;  \linebreak     {\tt gorosabel@stsci.edu, fruchter@stsci.edu,
   burud@stsci.edu, rhoads@stsci.edu}
 \and
 Th\"uringer Landessternwarte Tautenburg, D--07778 Tautenburg, Germany;
 {\tt klose@tls-tautenburg.de}
 \and
  Astrophysikalisches Institut, D--14482 Potsdam,
 Germany; {\tt lchristensen@aip.de}
 \and
  Department of Physics and Astronomy,  University of Aarhus, Ny
 Munkegade, DK--8000 \AA rhus C, Denmark; {\tt jfynbo@phys.au.dk}
 \and
  Astronomical Observatory, University of
 Copenhagen, Juliane Maries Vej 30, DK--2100 Copenhagen \O, Denmark; {\tt
   jens@astro.ku.dk, brian\_j@astro.ku.dk, holger@astro.ku.dk}
 \and
  Max-Planck-Institut f\"ur extraterrestrische Physik, D--85741 Garching,
 Germany; {\tt jcg@mpe.mpg.de}
 \and
 Department of Physical Sciences, University of Hertfordshire, College
 Lane, Hatfield, Herts AL10 9AB, UK; {\tt nrt@star.herts.ac.uk} 
 \and
  Department of Physics, University of
 Notre Dame, Notre Dame, IN 46556--5670, USA; {\tt sholland@nd.edu}
 \and
 European Southern Observatory, Casilla 19001, Santiago 19, Chile; {\tt
   ajaunsen@eso.org, pvreeswi@eso.org}
\and
  Real Instituto y Observatorio  de la Armada, Secci\'on de
 Astronom\'\i a, 11.110 San  Fernando-Naval (C\'adiz) Spain; \\
 {\tt josemari@alumni.nd.edu}
\and
 Osservatorio Astronomico di Trieste, Via G.B. Tiepolo 11, I--34131, Trieste,
 Italy; {\tt pian@tesre.bo.cnr.it}
 \and
 Dipartimento di Fisica, Universit\`a di Ferrara, Via Paradiso 12, I--44100
 Ferrara, Italy; {\tt frontera@fe.infn.it}
 \and
 Istituto  Tecnologie e  Studio Radiazioni  Extraterrestri,  CNR, Via
 Gobetti 101,  I--40129 Bologna, Italy;  \\ {\tt filippo@bo.iasf.cnr.it,
 masetti@bo.iasf.cnr.it, eliana@bo.iasf.cnr.it}
 \and
  University  of Amsterdam,  Kruislaan  403, NL--1098  SJ Amsterdam,  The
 Netherlands;                                                     {\tt
 lexk@science.uva.nl,evert@science.uva.nl,isabel@science.uva.nl,\\
 rwijers@science.uva.nl, edvdh@science.uva.nl}
\and
 NASA    MSFC,    SD--50,    Huntsville,    AL   35812,    USA;    {\tt
  kouveliotou@eagles.msfc.nasa.gov} }
\offprints{ \hbox{J. Gorosabel, e-mail:{\tt jgu@iaa.es}}}

\date{Received / Accepted }

\abstract{We report on multi-band ($UBVRIZJ_sK_s$) observations of the
  host galaxy  of the  April 18, 2000  gamma-ray burst.   The Spectral
  Energy  Distribution  (SED) is  analysed  by  fitting empirical  and
  synthetic spectral templates.  We find that: (i) the best SED fit is
  obtained with a starburst template, (ii) the photometric redshift is
  consistent with the spectroscopic redshift, (iii) the colours of the
  host are inconsistent  with an old stellar population,  and (iv) the
  global  extinction  is  constrained  to  be  in  the  range  $A_{\rm
  V}=0.12$--$0.61$ mag. The derived  global extinction agrees with the
  one  reported for  the  afterglow ($A_{\rm  V}  = 0.4$--$0.9$  mag),
  suggesting  a homogeneous  distribution of  the  interstellar medium
  (ISM)  in  the host  galaxy.   These  findings  are supplemented  by
  morphological information from Hubble Space Telescope (HST) imaging:
  the surface brightness profile is smooth, symmetric and compact with
  no underlying structures (like dust  lanes, spiral arms or disks). A
  natural  scenario which  accounts  of  all the  above  results is  a
  nuclear  starburst that harbours  a young  population of  stars from
  which  the  GRB  originated.     }

\maketitle

\section{Introduction}

Since 1997  $\sim$ 40 gamma-ray  burst (GRB) optical  afterglows (OAs)
have   been    discovered   (see    the   GRB   compilation    of   J.
Greiner\footnote{http://www.mpe.mpg.de/$\sim$jcg/grbgen.html}).    GRBs
generally occur  in subluminous  host galaxies with  redshifts ranging
from $z=0.1685$ (GRB~030329; Hjorth et al.  \cite{Hjor03}) to $z=4.50$
(GRB~000131; Andersen  et al.  \cite{Ande00}).  Most of  the GRB hosts
are subluminous and have  bluer optical/near-IR colours than the local
galaxies  or  the typical  galaxies  detected  by  the Infrared  Space
Observatory  (ISO) and the  Submillimeter Common-User  Bolometer Array
(SCUBA) (e.g.  Le Floc'h et al. \cite{LeFl03}).

The afterglow of GRB~000418  was discovered in the near-infrared (NIR)
2.5  days after the  gamma-ray event  (Klose et  al.  \cite{Klos00a}).
The optical counterpart was rather faint ($R=21.63$, $2.48$ days after
the GRB) in comparison to other afterglows (see Fig.~3 of Gorosabel et
al.   \cite{Goro02a}  for comparison  purposes).   The $R$-band  light
curve   decayed  as  $t^{-1.22}$   typical  of   OAs  (Klose   et  al.
\cite{Klos00b}), flattening off  at a level of $R \sim  24$ due to the
underlying host galaxy (Bloom et al. \cite{Bloo00}).  The afterglow is
one of the reddest ($R-K=4$) detected to date (see Fig.~2 of Gorosabel
et  al.  \cite{Goro02b}).   Klose et  al.   (\cite{Klos00b}) suggested
that the red  colour is caused by reddening due  to dust extinction in
the host  galaxy and they estimated  an extinction of  $A_{\rm V} \sim
0.9$ mag.  Berger  et al.  (\cite{Berg01}) found $A_{\rm  V} \sim 0.4$
mag for the OA.

HST/STIS observations performed on 4.17 UT June 2000 (46.76 days after
the GRB) revealed  that the OA occurred in a  very compact host galaxy
with  a  half-light  radius   of  $\sim  0\farcs13$  (Metzger  et  al.
\cite{Metz00}) corresponding to about 1 kpc.  The redshift of the host
galaxy was  determined to be  $z=1.118$ (Bloom et  al.  \cite{Bloo02},
\cite{Bloo03}). A preliminary  $BVRI$-band SED fitting analysis showed
that  the  host  galaxy  SED  can  be  fitted  with  starburst  galaxy
templates, but  not with an  evolved stellar population  (Gorosabel et
al.  \cite{Goro01}). This result  has been recently supported by Bloom
et  al.   (\cite{Bloo03}) who  based  on  the  optical emission  lines
suggest that the host is a  starburst galaxy, rather than a LINER or a
Seyfert 2 galaxy.  Bloom et al. (\cite{Bloo03}) estimate an unobscured
star formation  rate (SFR)  of $55 M_{\odot}$  yr$^{-1}$ based  on the
[\ion{O}{II}]    emission    line    diagnostic   method    (Kennicutt
\cite{Kenn92}).

Berger et al. (\cite{Berg03}) and Barnard et al.  (\cite{Barn03}) have
recently reported tentative detections of several GRB host galaxies in
the  sub-millimeter and radio  ranges, inferring  SFRs of  hundreds of
Solar masses  per year.  The  most significant detection was  from the
host  galaxy   of  GRB~000418  for  which  the   sub-mm  detection  is
significant at the  3.6$\sigma$ level and the radio  emission from the
host  again at more  than 3$\sigma$.   Berger et  al.  (\cite{Berg03})
also  detected another  faint radio  source only  1.4 arcsec  from the
GRB~000418  host.  This  source is  not seen  in the  optical  down to
R$>$27. Assuming  that the  sub-mm and radio  emission is due  to dust
heated by star formation Berger et al. (\cite{Berg03}) derived SFRs of
690$\pm$195  and 330$\pm$75  $M_{\odot}$ yr$^{-1}$  respectively, i.e.
much higher  than the SFR derived  from the optical  emission lines by
Bloom et al. (\cite{Bloo03}).

In  this paper  we present  a comprehensive  multicolour study  of the
GRB~000418 host galaxy, similar  to those performed for the GRB~000210
(Gorosabel et  al.  \cite{Goro03}) and GRB~990712  (Christensen et al.
\cite{Chri03}) host galaxies.  The aim of the analysis is to determine
the   properties   of   the   stellar   populations   dominating   the
optical/near-IR  light  from  the   host  galaxy  and  the  amount  of
extinction due to  dust in the interstellar medium  (ISM) of the host.
Other  multi-colour host  galaxies  studies to  date  (Sokolov et  al.
\cite{Soko01},   Chary  et  al.    \cite{Char02},  Gorosabel   et  al.
\cite{Goro01}) have been limited to a smaller number of bands.

Throughout,  the  assumed  cosmology  will  be  $\Omega_{\Lambda}  =  0.7$,
$\Omega_{M}  =  0.3$  and  $H_0=   65$  km  s$^{-1}$  Mpc$^{-1}$.   At  the
spectroscopic redshift of  the host galaxy ($z=1.118$), the  look back time
is 8.78 Gyr ($\approx$ 60\% of the present age) and the luminosity distance
is  8.17 Gpc.   The physical  transverse size  of one  arcsec  at $z=1.118$
corresponds to 8.83 kpc.

\section{Observations and photometry}
\label{Observations}

\begin{table*}
\begin{center}
\caption{Chronologically ordered optical and NIR observations carried
 out for the GRB~000418 host galaxy.}
\begin{tabular}{lcccr}
\hline
Telescope     &  Filter & Date UT &T$_{\rm exp}$&  Seeing \\
(+Instrument) &         &         &  (s)       &          \\
\hline
HST  (+STIS)   & 50CCD & $11.638 - 11.729/02/01$&   8$\times$640&  \#\\
2.5NOT (+ALFOSC) &$I$ &  $30.909 - 31.043/03/01$&  19$\times$300& 0.85\\
2.5NOT (+ALFOSC) &$V$ &  $31.067 - 31.135/03/01$&   6$\times$900& 0.95\\
2.5NOT (+ALFOSC) &$R$ &  $31.921/03 - 1.063/04/01$&  18$\times$600&1.10$^{\dagger}$\\
2.5NOT (+ALFOSC) &$B$ &  $1.064 - 1.189/04/01$&  8$\times$1200& 1.15\\
2.5NOT (+ALFOSC) &$Z$ &  $28.901 - 28.973/05/01$&  15$\times$300&0.85$^{\star}$\\
2.5NOT (+ALFOSC) &$Z$ &  $29.899-29.918/05/01$&   4$\times$300&1.20$^{\star}$\\
2.5NOT (+ALFOSC) &$Z$ &  $30.897 - 30.963/05/01$&  14$\times$300&1.10$^{\star}$\\
8.2VLT (+ISAAC)  &$K_s$& $15.000 - 15.054/06/01$&  30$\times$120& 1.20\\
8.2VLT (+ISAAC)  &$J_s$& $4.366 - 4.391/02/02$&  15$\times$120& 0.90\\
3.6ESO (+EFOSC2) &$U$  & $23.310 - 23.357/02/02$&   6$\times$600& 1.30\\
\hline
\multicolumn{5}{l}{\#  Not applicable.}\\
\multicolumn{5}{l}{$\dagger$  Through cirrus.}\\
\multicolumn{5}{l}{$\star$    The three epoch images were coadded resulting
in just a single Z-band magnitude.}\\
\hline
\label{table1}
\end{tabular}
\end{center}
\end{table*}

\begin{table*}
\begin{center}
\caption{Magnitudes of the host in  the STIS $CL$ and the ground-based
 $UBVRIZJ_sK_s$  bands.  Several  characteristics of  the  filters are
 displayed:  filter name  (1), effective  wavelength (2)  and bandpass
 width (3).  The  fourth column shows the measured  magnitudes (in the
 Vega  system  and  corrected  from  Galactic  reddening,  considering
 $E(B-V)=0.033$  given   by  Schlegel  et   al.   \cite{Sche98}).   To
 facilitate the calculation of the AB magnitudes, and consequently the
 flux  densities for each  band, the  AB offsets  are provided  in the
 fifth column. }
\begin{tabular}{lcccc}
\hline
Filter              & Effective       & Bandpass    &Magnitude &ABoff\\
name                & wavelength (\AA)& width (\AA) &          &     \\
\hline

$U$ (ESO\#640)       &   3711 &  166 & 23.54$\pm$0.30 &$ 0.82$\\
$B$                  &   4384 &  700 & 24.07$\pm$0.05 &$-0.13$\\
$V$                  &   5368 &  527 & 23.80$\pm$0.06 &$ 0.00$\\
STIS(50CCD)          &   6218 & 3538 & 23.76$\pm$0.10 &$ 0.21$\\
$R$                  &   6627 &  768 & 23.39$\pm$0.05 &$ 0.26$\\
$I$                  &   8007 &  778 & 22.79$\pm$0.05 &$ 0.46$\\
$Z$                  &   8940 &  993 & 22.46$\pm$0.10 &$ 0.55$\\
$J_s$ (ISAAC)        &  12499 &  958 & 22.27$\pm$0.10 &$ 0.94$\\
$K_s$ (ISAAC)        &  21638 & 1638 & 21.19$\pm$0.30 &$ 1.87$\\
\hline
\label{table2}
\end{tabular}
\end{center}
\end{table*}

\begin{figure}[t]
\begin{center}
  {\includegraphics[width=\hsize]{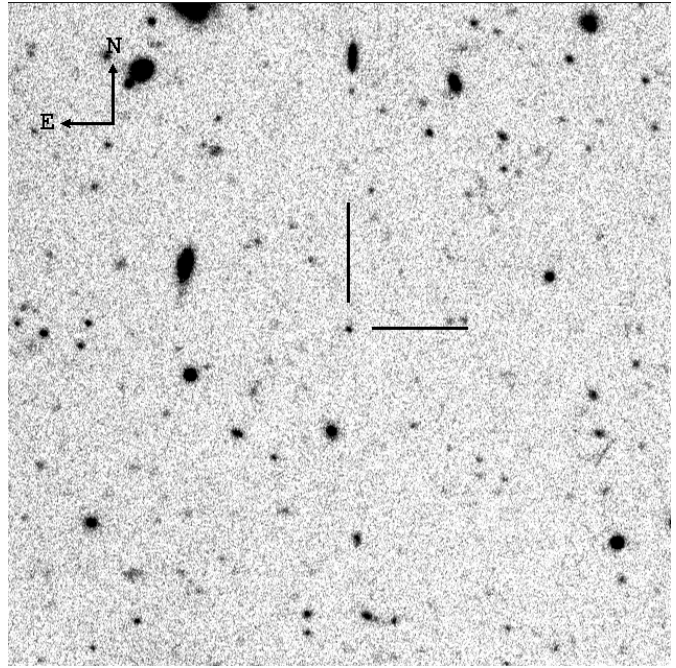}}
\caption{\label{2.5NOT}  The image  shows the  coadded  $V$-band image
 taken  with the  2.5NOT at  31.067$-$31.135 March  2001 UT.   The FOV
 covered by the image corresponds to $2\farcm0 \times 2\farcm0$. North
 is to the top and East to the left.}
\end{center}
\end{figure}

We have used ground-based and  space optical/NIR resources in order to
sample the host  galaxy SED in as many filters as  possible over a wide
spectral  range  (see   Table~\ref{table1}).   The  observations  were
performed 299--676 days after the gamma-ray event.  At these times the
contribution  of the  afterglow is  negligible and  does not  have any
significant impact on the derived host galaxy SED.

\subsection{Ground-based optical and near-infrared observations}

Given the compactness  of the GRB~000418 host galaxy,  its ground-based NIR
and optical profiles are consistent with that of field stars. This fact has
been checked by comparing the growth  curve (from 0.5 to 4 times their full
width half  maximum; FWHM  hereafter) of the  host and the  secondary stars
used for  photometric calibration.  Thus, for the  ground-based optical and
NIR pixel scales used in  the present study ($\ge 0\farcs148$/pix, achieved
with  ISAAC), the  GRB~000418 host  galaxy can  safely be  assumed to  be a
point-source.  Therefore,  considering  that  the  relative  photometry  is
independent  of  the aperture  radius,  the $UBVRIZJ_sK_s$-band  magnitudes
shown in  Table~\ref{table2} are based  on circular aperture  (PHOT running
under IRAF\footnote{IRAF  is distributed by the  National Optical Astronomy
  Observatories, which  is operated by the Association  of Universities for
  Research in Astronomy, Inc.   (AURA) under cooperative agreement with the
  National Science Foundation.}) with no aperture corrections.

$BVRIZ$-band  frames were  taken with  ALFOSC at  the 2.5-m  Nordic Optical
Telescope  (2.5NOT).  The  ALFOSC  detector is  a 2048$\times$2048  Thinned
Loral  CCD   providing  a   pixel  scale  of   $0\farcs189$/pix.   $U$-band
observations  were  carried  out  with  the 3.6-m  ESO  telescope  (3.6ESO)
equipped with  EFOSC2, covering a field  of view (FOV)  of $5\farcm5 \times
5\farcm5$.  These observations were carried out in 2$\times$2 binning mode,
providing a pixel scale of $0\farcs314$/pix.

The optical  data were  reduced in a  standard manner  (overscan, bias
subtraction,  and   division  by   a  normalised  flat   field).   The
$UBVRI$-band calibration was based  on the calibration given by Henden
(\cite{Hend00}).  The  $Z$-band calibration was  carried out observing
the spectro-photometric standard star Feige 66 (Oke \cite{Oke90}) with
the 2.5NOT(+ALFOSC) at an  airmass almost identical ($\Delta \sec(z) =
0.03$)  to the  GRB field.   Fig.~\ref{2.5NOT} shows  a  deep $V$-band
image of the GRB field taken with the 2.5NOT(+ALFOSC).

The NIR ($J_s$  and $K_s$-band) observations were acquired  with the UT1 of
the 8.2-m VLT (8.2VLT) equipped with ISAAC, allowing us to cover a $2\farcm
5 \times 2\farcm  5$ FOV with a pixel scale  of $0\farcs148$/pix.  In Table
\ref{table1}  we  provide  the  observing   log  of  our  optical  and  NIR
observations.  The  calibration was based  on observations of  the standard
stars  S301-D  ($J_s$  band)  and  S860-D  ($K_s$  band;  Persson  et  al.  
\cite{Pers98}). Due to the lack  of $J_s$-band calibration data for S301-D
we  assumed $J=J_s$.   Extinction coefficients  of  0.06 and  0.07 mag  per
airmass  (ESO Paranal)  where assumed  for $J_s$  and $K_s$,  respectively. 
Given that  the airmass difference between  the GRB field  and the standard
was $\Delta \sec(z) = 0.4$ in  both $J_s$ and $K_s$, the introduced airmass
correction is well below our measurement error on the magnitude of the host
galaxy.

\begin{figure*}[t]
\begin{center}
{\includegraphics[width=8.6cm]{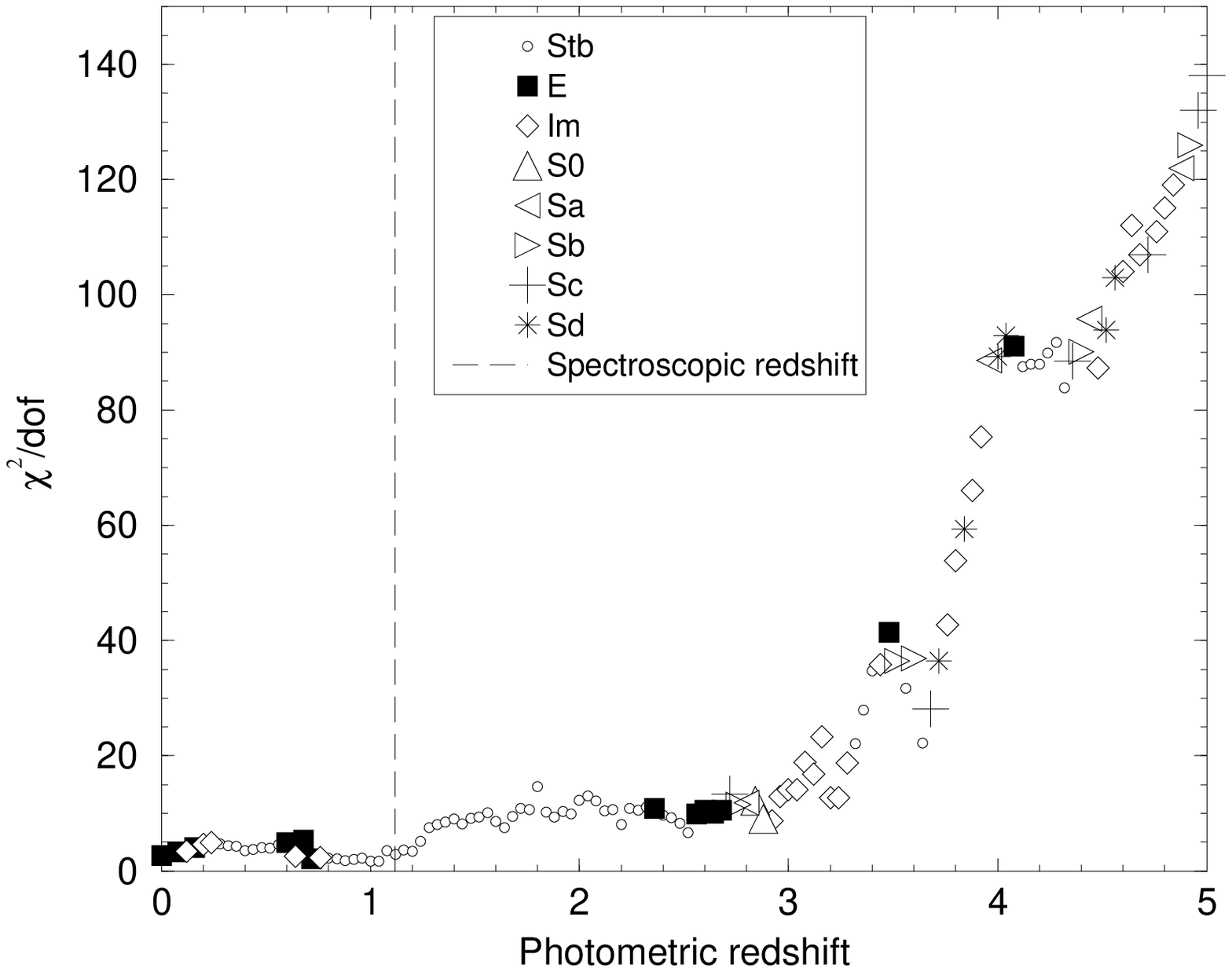}}
{\includegraphics[width=8.4cm]{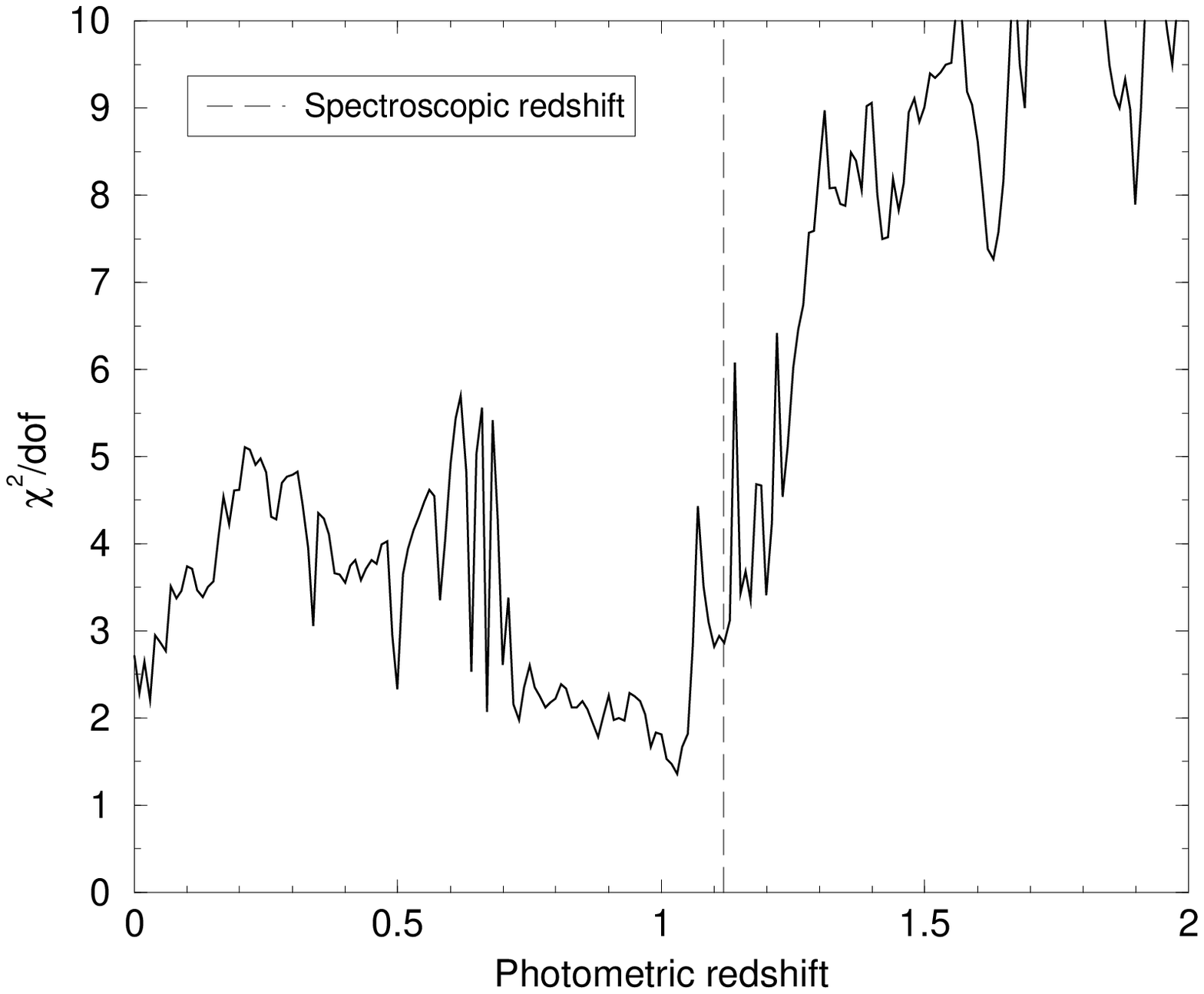}}
 \caption{\label{chi2} {\em Left panel:} The global mapping of the  fitted
   SED $\chi^2/dof$  when the  redshifts of the  templates are varied  in a
   broad  range ($0  <  z <5$).   The  dotted vertical  line indicates  the
   spectroscopic redshift  of the  host galaxy ($z=1.118$).   The different
   symbols indicated the  best fitted template type for  each redshift.  As
   it is  shown, the  spectroscopic redshift is  consistent with  the broad
   minimum  ($0.65 <  z <  1.20$) centred  at $z=1.029$.   This  minimum is
   governed by Stb  galaxies.  Other local minima (for  instance the one at
   $z  \sim 0$  achieved with  E  and Im  galaxies) show  larger values  of
   $\chi^2/dof$.  {\em  Right panel:} The plot  shows a blow  up around the
   absolute  minimum  with  no  distinction  among  templates  types.   The
   evolution of $\chi^2/dof$ rejects  the photometric redshifts not located
   at $0.65 < z < 1.2$, specially  the ones at $z > 1.5 $ where $\chi^2/dof
   > 9$.  {\em  General:} The plot assumed  a Sc86 IMF,  a Sea79 extinction
   law, and Solar metallicity (SOL-Sc86-Sea79 subfamily of templates).}
\end{center}
\end{figure*}

The photometric  calibration has been  tested performing photometry  of the
2MASS star at RA, DEC (J2000) = 12:25:14.47, 20:05:49.8 which is present on
the ISAAC images taken for the  host. Using the S301-D star we derived for
this  source $J_s  =  15.51 \pm  0.01$  and $K_s=  14.87  \pm 0.01$.   Both
magnitudes  agree with  the  ones  given in  the  2MASS Second  Incremental
Release Point Source  Catalog ($J_s = 15.47 \pm 0.06$ and  $K_s = 14.69 \pm
0.10$).

\subsection{HST observations}

The HST  observations were  carried out under  the ToO  programme 8189
(Principal  Investigator,  Fruchter   et  al.   \cite{Fruch99}).   The
observations were  performed at an epoch  299 days after  the GRB (see
Table  \ref{table1}) where  the  OA contribution  is negligible.   The
observations  were performed  with  STIS,  which yields  a  gain of  1
$e^{-}/ADU$,  a physical  pixel  scale  of $0\farcs05$  and  a FOV  of
$52^{\prime \prime} \times  52^{\prime \prime}$ (STScI \cite{Stsc00}).
The  data were  obtained  with the  50CCD  or $CL$  filter.  The  STIS
reduction pipeline {\tt calstis} of IRAF was used to calibrate the raw
data.  The  raw images were reduced following  normal procedures (bias
and dark subtraction,  and division by a normalised  flat field).  The
individual images were  combined using the {\tt drizzle}  task of IRAF
(Fruchter \& Hook  \cite{Fruch02}).  In the drizzling of  the data the
parameters  {\tt pixfrac=0.6}  and {\tt  scale=0.5} were  used.  These
values yield  an output  grid of  2k $\times$ 2k  pixels with  a pixel
scale of  $0\farcs0254$/pix.  We performed aperture  photometry of the
host  with a  radius  of 50  drizzled  pixels.  The  count rate  ($C$,
measured  in $ADU$/s) was  converted into  the $CL$-band  AB magnitude
following:   $m^{\rm   AB}_{CL}=26.386  -   2.5   \log   C  $   (STScI
\cite{Stsc00}).

In  order to fit  the observed  Vega system  magnitudes ($m$)  using a
given template (either empirical or synthetic), the values of $m$ were
converted  into  flux densities  ($f_{\nu}$)  using the  corresponding
offsets to the AB magnitude  system (Oke \cite{Oke90}).  The AB offset
is defined as ABoff$= m_{\rm AB} - m$, where $m_{\rm AB} = -2.5 \times
\log f_{\nu}  - 48.60$ ($f_{\nu}$  measured in erg  s$^{-1}$ cm$^{-2}$
Hz$^{-1}$) is the  magnitude in the AB system.  The  AB offsets of our
nine bands  were derived convolving  the Vega spectrum taken  from the
GISSEL98  (Bruzual \& Charlot  \cite{Bruz93}) library  ($\alpha$ Lyrae
$m=0$  in  all bands  by  definition)  with the  $UBVRIZJ_sK_sCL$-band
filters    and    the    corresponding    CCD    efficiency    curves.
Table~\ref{table2} displays  the AB offsets  and effective wavelengths
of the nine bands used to  construct the SED.  Prior to performing the
SED  fit,  the   derived  $UBVRIZJ_sK_sCL$-band  flux  densities  were
dereddened of  the Galactic  extinction in the  direction of  the host
($E(B-V) = 0.033$; Schlegel et  al.  \cite{Sche98}).  The use of other
Galactic  extinction maps  does not  affect the  final results  of our
analysis (Dickey \& Lockmann \cite{Dick90} give $E(B-V) = 0.057$).

\section{Methodology}
\label{method}

\subsection{Construction of synthetic SED templates}
\label{synseds}

The applied synthetic SED fitting technique  is the same as the one applied
to the host  galaxy of GRB~000210 (Gorosabel et  al.  \cite{Goro03}) and is
based on Hyperz\footnote{ http://webast.ast.obs-mip.fr/hyperz/} (Bolzonella
et al.   \cite{Bolz00}).  Eight  synthetic spectral types  were constructed
representing Starburst  galaxies (Stb), Ellipticals  (E), Lenticulars (S0),
Spirals  (Sa,  Sb,  Sc and  Sd)  and  Irregular  galaxies (Im).   The  time
evolution of the SFR for all  galaxy types is represented by an exponential
model, i.e.   SFR $\propto  \exp(-t/\tau)$, in which  $\tau$ ranges  from 0
(Stb) to 30 Gyr  (Sd).  The SFR of Stb is modeled  by an exponential decay
in the  limit $\tau  \rightarrow 0$, in  other words an  instantaneous star
burst. The early  type galaxy spectra (E, S0) are  represented by values of
$\tau$ between 1  and 2 Gyr.  The  SFR of Im galaxies are  represented by a
constant SFR ($\tau \rightarrow \infty$).

The  impact  of  the  assumed  Initial Mass  Function  (IMF)  and  the
extinction  law have been  tested.  Three  IMFs were  assumed, namely:
Miller \&  Scalo (\cite{Mill79}; MiSc79),  Scalo (\cite{Scal86}; Sc86)
and    Salpeter   (\cite{Salp55};    Sp55).    Christensen    et   al.
(\cite{Chri03})  show that the  GRB~990712 host  galaxy SED  is better
reproduced  with a  Sp55 IMF.   However,  the same  method applied  to
GRB~000210 did  not show any preference  in the IMF  (Gorosabel et al.
\cite{Goro03}).  According to Bolzonella  et al.   (\cite{Bolz00}) the
Sp55 IMF produces an excess of  bright blue stars producing an UV flux
excess.  On  the other  hand, the  Sc86 IMF creates  an excess  in the
number of solar  mass stars, making the spectrum  too red to reproduce
the photometric  points.  Intensive photometric  redshift studies have
shown that the MiSc79 IMF is a good compromise between both tendencies
(Bolzonella et al.  \cite{Bolz02}).

We  have  also   tried  to  infer  information  on   the  host  galaxy
metallicity.   The major  effect of  increasing the  amount  of metals
contained in the host is to produce redder colours, hence altering the
extinction  determination.  The  impact  of the  metallicity has  been
tested  comparing the results  obtained when  solar metallicity  ($Z =
Z_{\odot} \simeq  0.02$; SOL)  and evolving metallicity  templates are
used.  The metallicity evolution  is based on the so-called closed-box
model,  which  considers the  ejection  of  heavy  elements from  each
generation  of  stars assuming  an  instantaneous  recycling of  heavy
elements.  The two  template groups (SOL and EVOL)  have been combined
with  the above  mentioned three  IMFs, constituting  6  SED templates
families (SOL-MiSc79,  SOL-Sc86, SOL-Sp55, EVOL-MiSc79,  EVOL-Sc86 and
EVOL-Sp55).

\begin{figure*}[t]
\begin{center}
 {\includegraphics[width=8.7cm]{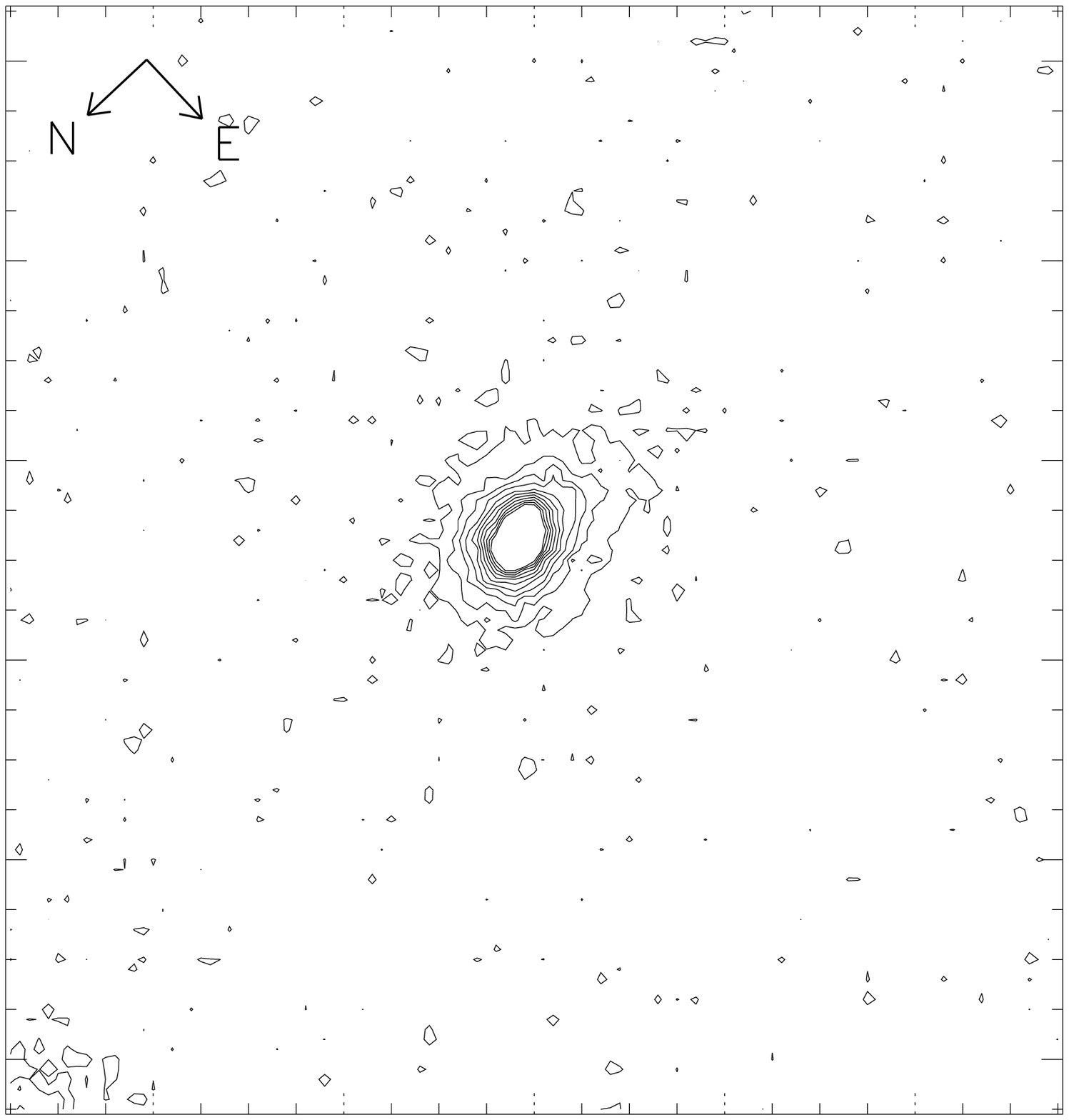}}
 {\includegraphics[width=8.7cm]{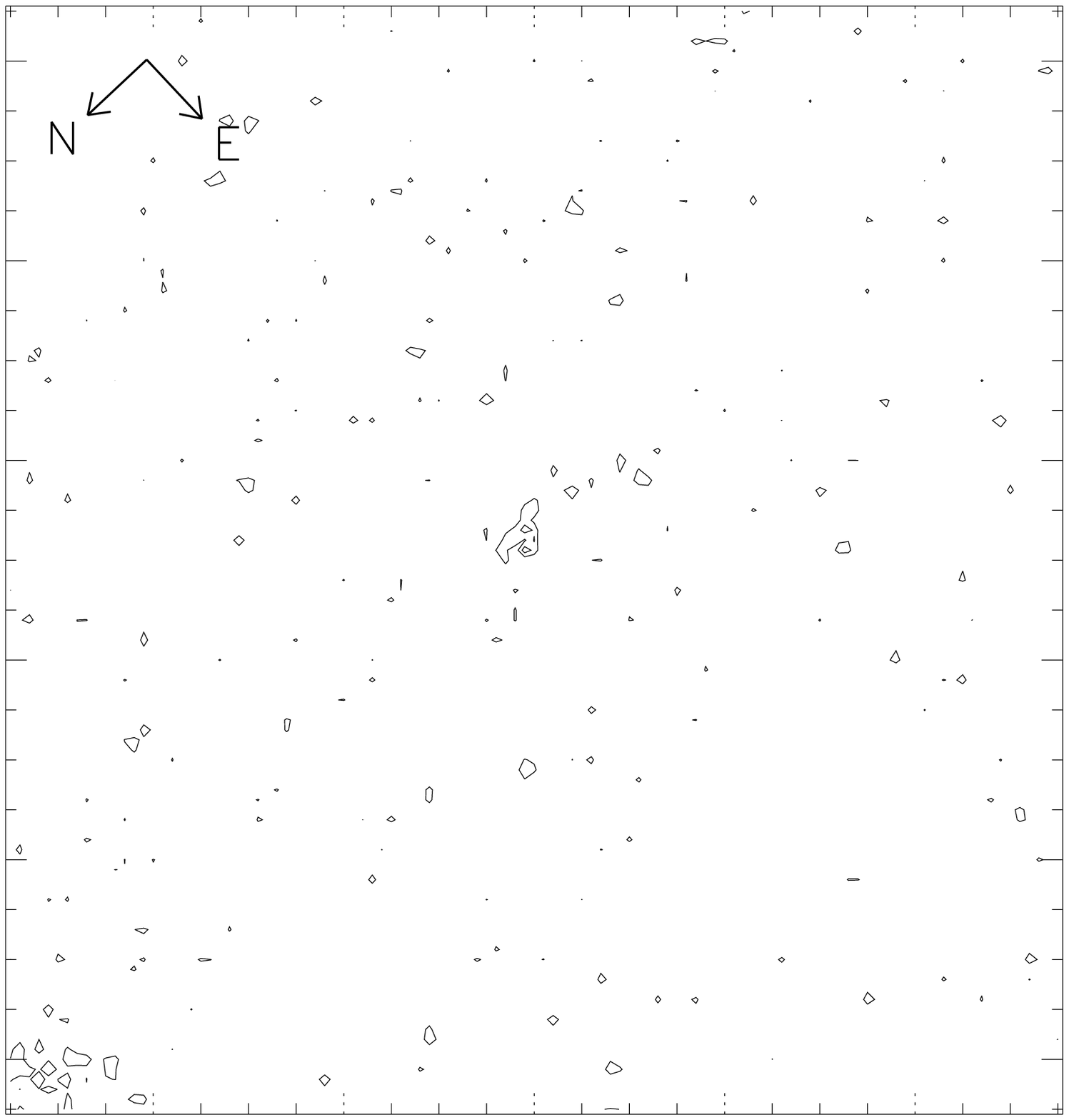}}
\caption{\label{stis}  The left  panel  shows  a contour
  plot of  the GRB~000418  host galaxy seen  with STIS ($CL$-band  filter). 
  The observations were carried out  299 days after the gamma-ray event, so
  no contribution from  the OA is expected.  The  right panel contours show
  the residuals of  the host when the host  galaxy surface brightness model
  is  subtracted (see  Sect.~\ref{ellipse} for  further  information).  The
  contours  of both figures  scale linearly  with the  detection confidence
  level (from $2\sigma$ to $30\sigma$, with a step of 3$\sigma$). As can be
  seen in the  right plot, the residuals consist of  a few pixels $2\sigma$
  above the background.   They are concentrated in the  host nucleus, where
  the PSF of  STIS might affect the surface  photometry.  Thus, we conclude
  that (perhaps with the exception of the central region within a radius of
  $0\farcs075$) there  are no reliable underlying  structures associated to
  the  host  galaxy (like  dust  lanes, spiral  arms  or  disks) above  the
  detection  threshold of  our STIS  images.  The  FOV is  $2\farcs8 \times
  2\farcs8$.  The arrows indicate the orientation of the images.}
\end{center}
\end{figure*}

\begin{table*}
\begin{center}
\caption{The table displays the parameters of the best host galaxy SED fit
  when several IMFs and extinction  laws, indicated in the first and second
  columns,  are adopted.   The rest  of  the columns  display the  inferred
  parameters under  the assumed IMF  and extinction law.  The  third column
  provides the  confidence of  the best fit  (given by  $\chi^2/dof$, being
  $dof=8$).  The  derived photometric redshift  is displayed in  the fourth
  column (and the  corresponding 68\% and 99\% percentile  errors).  In the
  fifth and  sixth columns the template  family of the best  fitted SED and
  the age of the stellar population are given.  The seventh column displays
  the derived value of the  host galaxy extinction $A_{\rm V}$.  The eighth
  column displays the derived rest  frame absolute B-band magnitude, $M_B$. 
  The ninth column gives the Luminosity of the host in units of $L^{\star}$
  (Schechter \cite{Sche76}).   The Last column  displays SFR$_{UV}$ derived
  from  the 2800  \AA~  flux once  it  is corrected  for the  corresponding
  reddening  (see   Sect.~\ref{sfr}  for  a  detailed   discussion  on  the
  SFR$_{UV}$ estimation).   The table is divided in  three sub-tables.  The
  upper  sub-table displays the  derived parameters  when a  constant solar
  metallicity is assumed.  The middle one assumes a metallicity evolving in
  time.  The  lower sub-table  displays the results  when no  extinction is
  assumed. }
\begin{tabular}{lccccccccc}
\hline
IMF & Extinction Law & $\chi^{2}/dof$& Photometric Redshift & Template &  Age  &$A_{\rm V}$& $M_B$&$L/L^{\star}$&SFR$_{UV}$ \\
    &                &  ($dof=8$)    &$z^{+ p68\%,~ p99\%}_{- p68\%,~ p99\%}$&&(Gyr)&         & & &$M_{\odot}$ yr$^{-1}$ \\
\hline
\hline
\multicolumn{10}{c}{ Constant metallicity ($Z = Z_{\odot} \simeq 0.02$)}\\
\hline
\hline
Sp55  &Cal00&$0.724$&$1.005^{+0.022, 0.057}_{-0.004, 0.281}$&Stb&0.004&1.38&-20.41&0.84& 60.4$\pm$27.3\\
MiSc79&Cal00&$0.835$&$1.004^{+0.004, 0.055}_{-0.010, 0.254}$&Stb&0.004&1.47&-20.38&0.82& 70.1$\pm$33.7\\
Sc86  &Cal00&$0.739$&$1.004^{+0.027, 0.057}_{-0.006, 0.250}$&Stb&0.004&1.35&-20.40&0.83& 57.5$\pm$25.4\\
Sp55  &Pre84&$0.723$&$1.001^{+0.004, 0.052}_{-0.011, 0.017}$&E  &0.181&0.42&-20.39&0.82& 14.2$\pm$ 1.0\\
MiSc79&Pre84&$0.750$&$1.001^{+0.004, 0.052}_{-0.010, 0.052}$&Im &0.128&0.39&-20.39&0.82& 13.4$\pm$ 0.9\\
Sc86  &Pre84&$0.673$&$1.020^{+0.009, 0.033}_{-0.004, 0.075}$&E  &0.128&0.36&-20.44&0.86& 12.6$\pm$ 0.8\\
Sp55  &Fit86&$1.349$&$1.020^{+0.012, 0.032}_{-0.005, 0.068}$&Stb&0.053&0.18&-20.43&0.86&  8.5$\pm$ 0.3\\
MiSc79&Fit86&$1.395$&$1.017^{+0.015, 0.035}_{-0.003, 0.069}$&Stb&0.053&0.18&-20.42&0.85&  8.5$\pm$ 0.3\\
Sc86  &Fit86&$1.319$&$1.029^{+0.005, 0.024}_{-0.014, 0.069}$&Stb&0.053&0.12&-20.46&0.88&  7.6$\pm$ 0.3\\
Sp55  &Sea79&$1.431$&$1.031^{+0.003, 0.022}_{-0.018, 0.079}$&Stb&0.064&0.12&-20.45&0.87&  7.6$\pm$ 0.3\\
MiSc79&Sea79&$1.472$&$1.019^{+0.014, 0.033}_{-0.006, 0.072}$&Stb&0.053&0.15&-20.43&0.86&  8.1$\pm$ 0.3\\
Sc86  &Sea79&$1.358$&$1.029^{+0.005, 0.025}_{-0.012, 0.069}$&Stb&0.053&0.12&-20.46&0.88&  7.6$\pm$ 0.3\\
\hline
\hline
\multicolumn{10}{c}{ Evolving metallicity }\\
\hline
\hline
Sp55  &Cal00&$0.724$&$1.005^{+0.022, 0.055}_{-0.022, 0.251}$&Stb&0.004&1.38&-20.41&0.84& 60.4$\pm$27.3\\
MiSc79&Cal00&$0.835$&$1.004^{+0.004, 0.055}_{-0.022, 0.253}$&Stb&0.004&1.47&-20.38&0.82& 70.1$\pm$33.7\\
Sc86  &Cal00&$0.739$&$1.004^{+0.027, 0.057}_{-0.006, 0.250}$&Stb&0.004&1.38&-20.40&0.83& 60.4$\pm$27.3\\
Sp55  &Pre84&$0.668$&$1.019^{+0.013, 0.040}_{-0.009, 0.323}$&E  &0.181&0.36&-20.46&0.88& 12.6$\pm$ 0.8\\
MiSc79&Pre84&$0.596$&$1.017^{+0.015, 0.043}_{-0.005, 0.320}$&E  &0.128&0.36&-20.46&0.88& 12.6$\pm$ 0.8\\
Sc86  &Pre84&$0.507$&$1.016^{+0.016, 0.043}_{-0.005, 0.061}$&E  &0.128&0.33&-20.46&0.88& 11.9$\pm$ 0.7\\
Sp55  &Fit86&$1.349$&$1.020^{+0.012, 0.032}_{-0.005, 0.068}$&Stb&0.053&0.18&-20.43&0.86&  8.5$\pm$ 0.3\\
MiSc79&Fit86&$1.395$&$1.013^{+0.015, 0.036}_{-0.004, 0.069}$&Stb&0.053&0.18&-20.42&0.85&  8.5$\pm$ 0.3\\
Sc86  &Fit86&$1.319$&$1.029^{+0.015, 0.044}_{-0.014, 0.069}$&Stb&0.053&0.12&-20.46&0.88&  7.6$\pm$ 0.3\\
Sp55  &Sea79&$1.431$&$1.031^{+0.003, 0.022}_{-0.018, 0.079}$&Stb&0.064&0.12&-20.45&0.87&  7.6$\pm$ 0.3\\
MiSc79&Sea79&$1.472$&$1.019^{+0.003, 0.014}_{-0.006, 0.072}$&Stb&0.053&0.15&-20.43&0.86&  8.1$\pm$ 0.3\\
Sc86  &Sea79&$1.358$&$1.029^{+0.005, 0.025}_{-0.012, 0.069}$&Stb&0.053&0.12&-20.46&0.88&  7.6$\pm$ 0.3\\
\hline
\hline
\multicolumn{10}{c}{ No extinction }\\
\hline
\hline
Sp55  &  -- &$1.559$&$1.011^{+0.002, 0.040}_{-0.002, 0.007}$&Stb&0.091&0.0 &-20.42&0.85&  6.2$\pm$0.2\\
MiSc79&  -- &$1.659$&$1.011^{+0.002, 0.040}_{-0.002, 0.009}$&Stb&0.091&0.0 &-20.41&0.84&  6.2$\pm$0.2\\
Sc86  &  -- &$1.646$&$1.029^{+0.003, 0.023}_{-0.008, 0.034}$&Stb&0.064&0.0 &-20.47&0.89&  6.2$\pm$0.2\\
\hline
\label{synfitresults}
\end{tabular}
\end{center}
\end{table*}

\begin{table}
\begin{center}
\caption{Results obtained  based on the empirical  templates by Kinney
  et  al.  (1996).   The templates  have  been ordered  according to  their
  colour, bluer on the  top (Stb1) and redder on the bottom  (B). As it can
  be seen there is a clear  correlation between the goodness of the fit and
  the colour of the template: the  bluer the colour the lower $\chi^2/dof$. 
  While  Stb type galaxies  (Stb1, Stb2,  Stb3, Stb4,  Stb5 and  Stb6) have
  their extinction fixed by the template definition, for the quiescent ones
  (Sc, Sb, Sa,  S0, E and B)  the extinction can take any  arbitrary value. 
  In  all these early  and mid  type cases  the blue  SED predicted  by our
  photometric data is  reproduced more closely by the  bluest fit possible. 
  Introduction of  dust reddening for  Sc, Sb, Sa,  S0, E and B  would just
  worsen the fit.  This is why in the last column of the table $E(B-V)=0.0$
  for the quiescent galaxies.}
\begin{tabular}{lccr}
\hline
Galaxy   &$\chi^2/dof$&$E(B-V)$\\
Template &($dof=8$)   &        \\
\hline
Stb1    &$  5.92$    &$0.05\pm0.05^{\dagger}$\\
Stb2    &$  4.81$    &$0.15\pm0.05^{\dagger}$\\
Stb3    &$ 23.71$    &$0.30\pm0.05^{\dagger}$\\
Stb4    &$ 37.00$    &$0.45\pm0.05^{\dagger}$\\
Stb5    &$ 48.29$    &$0.55\pm0.05^{\dagger}$\\
Stb6    &$ 70.51$    &$0.65\pm0.05^{\dagger}$\\
Sc      &$ 66.23$    &$0.0$ \\
Sb      &$206.01$    &$0.0$ \\
Sa      &$225.86$    &$0.0$ \\
S0      &$249.38$    &$0.0$ \\
E       &$255.08$    &$0.0$ \\
B       &$255.65$    &$0.0$ \\
\hline
\multicolumn{3}{l}{$\dagger$ Extinction fixed by the template definition.}\\
\hline
\label{empfitresults}
\end{tabular}
\end{center}
\end{table}

\begin{figure*}[t]
\begin{center}
 {\includegraphics[width=16cm]{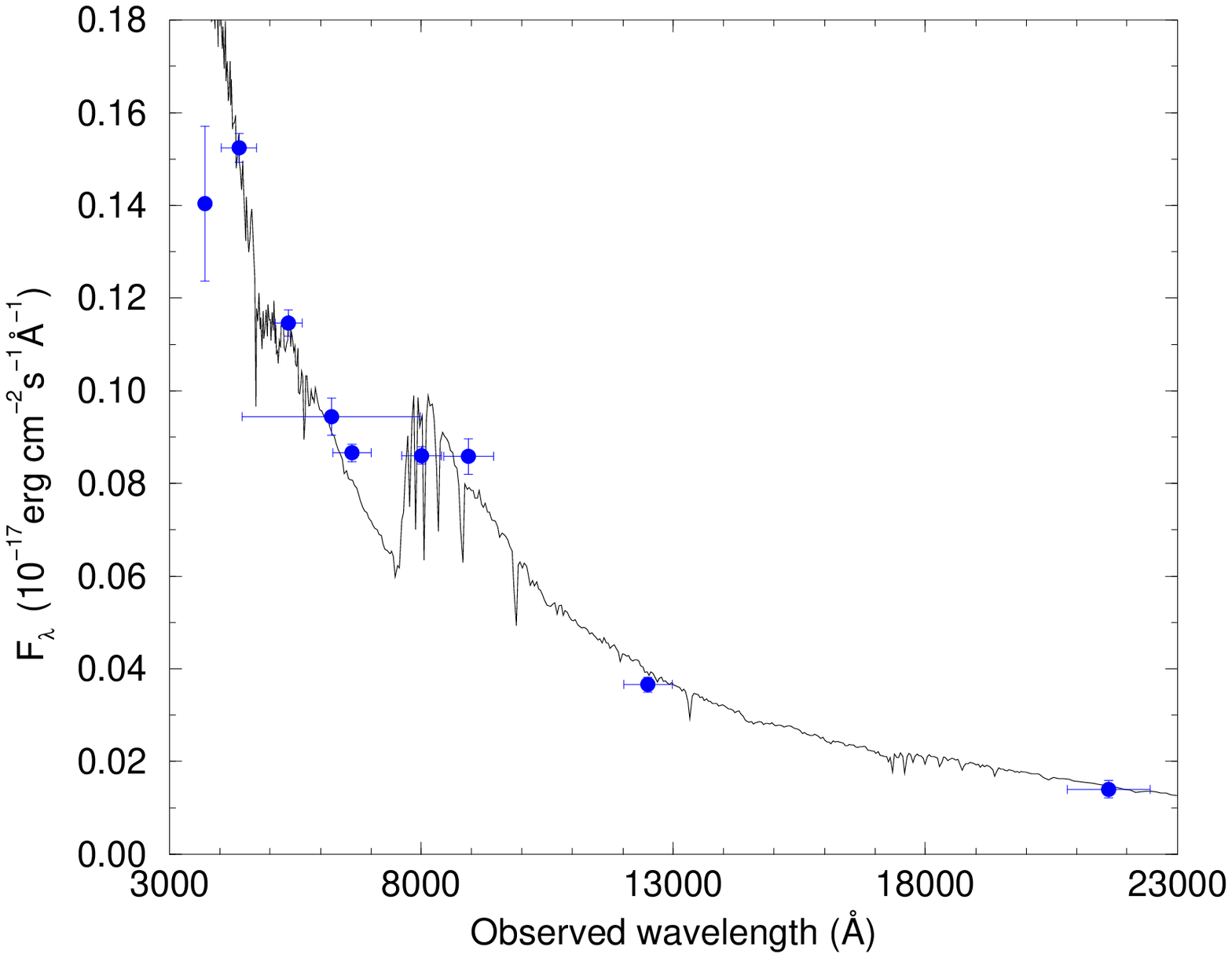}}
\caption{\label{418sed}  The  points show  the  measured  flux in  the
  $UBVRIZJ_sK_sCL$  bands for  the  GRB~000418 host  galaxy, once  the
  Galactic dereddening is  introduced ($E(B-V)=0.033$, Schlegel et al.
  \cite{Sche98}).  The  solid curve represent  the best SED  fitted to
  the photometric points ($\chi^2/dof$  = $1.358$), corresponding to a
  starburst  synthetic template  at a  redshift of  $z =  1.029$.  The
  plots assumes  a Sc86 IMF.  The  derived value of  the starburst age
  corresponds to  $0.053$ Gyr.  The  extinction law used  to construct
  the plot is given by Sea79.  The fit is consistent with a low global
  extinction in  the host ($A^{global}_{\rm  V} \sim 0.12$).   The SED
  shows   an  appreciable   $\sim  4000   \times   (1+z)$~\AA  ~break,
  approximately at the $Z$-band filter wavelength.}
\end{center}
\end{figure*}

Furthermore, the  effect of considering different extinction  laws has been
studied.  For each of the 6 families 4 extinction laws have been taken into
account for the determination of  the photometric redshift and the physical
conditions of  the host, namely:  Calzetti et al.   (\cite{Calz00}), Seaton
(\cite{Seat79}),   Fitzpatrick  (\cite{Fitz86}),  and   Pr\'evot  et   al.  
(\cite{Prev84}). The extinction laws  will be abbreviated as, Cal00, Sea79,
Fit86 and Pre84. Each of  these extinction laws specifies the dependence of
the  extinction with  frequency and  is  the result  of different  physical
conditions in the  interstellar space in the hosts.  Thus, Sea79, Fit86 and
Pre84, are  representative of  the Milky Way  (MW), Large  Magellanic Cloud
(LMC) and the  Small Magellanic Cloud (SMC) extinction  laws, respectively. 
The Cal00 extinction law is suitable for starburst regions.

Due to computational  limitations, the fitting strategy was  divided in two
steps.   As  a  first stage,  the  redshifts  and  the extinctions  of  the
templates were varied  in a broad range ($0  < z <5$, $0 < A_{\rm  V} < 5$;
see  left  panel   of  Fig.~\ref{chi2})  in  order  to   map  globally  the
$\chi^{2}/dof$ evolution.  The used  steps were $\Delta~z=0.05$ and $\Delta
A_{\rm V}=0.05$, respectively.  Once a deep broad minimum of $\chi^{2}/dof$
was localized around $z \sim  1$, the accurate fit parameters (displayed in
Table  \ref{synfitresults}) were  determined  with a  thinner  grid of  SED
templates  ($\Delta~z=0.01$, $\Delta  A_{\rm V}=0.03$;  see right  panel of
Fig.~\ref{chi2}).  In  this second stage,  $z$ and $A_{\rm V}$  were ranged
around the $\chi^{2}/dof$ minimum ($0 < z < 2$ and $0 < A_{\rm V} < 3$).

\subsection{Comparison with empirical SED templates}
\label{obsseds}

We have complemented the analysis  carried out with the synthetic templates
by  using the  12 empirical  templates  by Kinney  et al.   (\cite{Kinn96};
hereafter K96).   Those templates reproduce 6 SEDs  of quiescent (Bulge(B),
E, S0, Sa,  Sb and Sc) and  6 of starburst galaxies.  The  templates of the
quiescent  galaxies are  built  according to  morphological  type, and  the
starburst galaxies according to colour excess (Stb1, Stb2, Stb3, Stb4, Stb5
and Stb6; see K96).  The starburst galaxy SEDs are represented by a grid of
templates constructed increasing $E(B-V)$ from  0.05 to 0.65 with a step of
0.1 (see column 3 of Table \ref{empfitresults}).

The  early-type empirical  templates (B,  E and  S0) have  the reddest
colours (dominated  by evolved stellar populations) with  a large flux
density increase (in the  $F_{\lambda}$ representation) from the UV to
the optical wavelengths, specially at the ~4000~\AA~ break. The Sa and
Sb galaxies tend to be slightly bluer.  For all the early and mid type
templates  (B,  E, S0,  Sa  and  Sb)  the flux  density  $F_{\lambda}$
increases with lambda.   The Sc template is in  the transition towards
the  blue  starburst  galaxies  and  can be  approximated  by  a  flat
spectrum.   The  starburst templates  show  spectra  that rise  toward
shorter  wavelengths.   They become  increasingly  bluer, as  $E(B-V)$
decreases progressively from Stb6 to Stb1.

The wavelength  coverage of  the K96 templates  (1000-10000 \AA)  have been
extended  to the  near-IR by  means of  Bruzual \&  Charlot (\cite{Bruz93})
spectra with parameters selected to match the K96 spectra at $z=0$.  In any
case, for  the redshift  of the host  ($z=1.118$) the extrapolation  is not
crucial because  only the  $K_s$ band falls beyond  the restframe  IR limit
(10000 \AA) of the K96 templates.

The  fits have  been  performed fixing  the  redshift of  the  host at  the
spectroscopic redshift.  For the  quiescent galaxy templates the extinction
law given by Sea79 has been  assumed, leaving $E(B-V)$ as a free parameter. 
For the  starburst galaxies the used  extinction law was the  one of Cal00,
fixing $E(B-V)$ according to the  definition of the corresponding template. 
Table \ref{empfitresults} shows the results obtained for each template.

\section{Results}
\label{results}

\subsection{Underlying morphological structure}
\label{ellipse}

We have built  a model of the galaxy based on  ellipses of constant surface
brightness.  Once  the model  has been constructed  it has  been subtracted
from the  observed surface  brightness in order  to detect  residuals which
could reveal the presence of underlying structures (like dust lanes, spiral
arms  or  disks). This  method  was  already  successfully applied  to  the
HST/STIS data  taken for the  GRB~980703 host galaxy,  suggesting subjacent
structures or/and an irregular  morphology (Holland et al.\ \cite{Holl01}). 
Similar techniques  also found significant  structure in the inner  part of
AGNs, supporting the scenario of central supermassive black holes (Xilouris
\& Papadakis \cite{Xilo02}).

The mentioned ellipses of  constant surface brightness where obtained using
the {\tt isophote}  package in IRAF. The surface  photometry is analysed by
the task {\tt  ellipse}, which allows the construction  of the galaxy model
by means of the {\tt bmodel}  IRAF task.  The determination of the residual
image  was  performed  in  two  steps.  First  the  constructed  model  was
subtracted  from the  observed surface  brightness providing  a preliminary
residual.  This residual was used to make a mask that excluded other nearby
objects from the model.  If the  pixel value in the residual image deviates
by more than  2$\sigma$ from the input image, the pixel  is flagged and not
used in a new run of {\tt ellipse} and {\tt bmodel}.

The left  panel of Fig.~\ref{stis}  shows the coadded  STIS 50CCD-band
image of the  host galaxy. The right panel  shows the subtracted image
where only a  few pixels are more than  $2\sigma$ above the background
noise level.  We do not  detect any underlying structure  further than
$0\farcs075$ (0.66 kpc at $z=1.118$) from the center of the host.

A  detailed  morphological study  of  the  GRB~000418  host radial  profile
(taking into account  the effects of the HST/STIS PSF)  is beyond the scope
of  the   present  study.   Such   analysis  can  be  found   in  Vreeswijk
(\cite{Vree02}), who report  that both a de Vaucouleurs  and an exponential
profile provide reasonable fit to the data.

\subsection{Information derived from the SED fits}
Fig.~\ref{418sed}   shows  the  SED   of  the   GRB~000418  host   galaxy.  
Table~\ref{synfitresults}  displays the  result of  our SED  fits  based on
synthetic templates. Table  ~\ref{empfitresults} displays the complementary
results obtained when the empirical K96 templates are fitted.

The discussion on  the impact of the mentioned SED  fitting methods and the
corresponding assumed parameters is split as follows:

\subsubsection{Galaxy type}
\label{type}

According to Table~\ref{synfitresults} most  of the IMF and extinction
combinations favour  a starburst galaxy.  The exception  occurs when a
Pre84  extinction law  is assumed.   In most  of those  cases  a young
Elliptical galaxy template is the  best fitted SED, with a stellar age
0.128--0.181 Gyr.   In one case (SOL-MiSc79) the  Pre84 extinction law
is consistent with a young (0.128 Gyr) Im galaxy template.

In the first stages of the SED evolution, when the galaxy is dominated
by a young blue stellar  population (typically younger than $0.1$ Gyr)
all the synthetic templates  resemble each other.  The templates start
to differentiate at ages $>0.1$  Gyr, when the 4000~\AA~ break becomes
progressively prominent.   Thus, the  ages inferred for  the E  and Im
templates are  close to that evolutionary transition  present at $\sim
0.1$~Gyr, when  their SEDs are  still similar to starbursts.   The low
age derived  for the  E and  Im templates is  just an  indication that
undergoing  active   stellar  formation   is  present  in   the  host,
independently of  the type assigned  to each template  (the difference
among the  young templates is more  due to a  template name definition
than  a real  physical reason).   So, only  based on  the  analysis of
synthetic templates we  can not completely rule out  that the host SED
is not a young  E or an Im galaxy.  However, even  for these few cases
(which   represent   22~\%  of   the   SED   solutions  displayed   in
Table~\ref{synfitresults}), we  can claim  that the galaxy  harbours a
prominent  blue stellar  formation activity.   Therefore, it  is clear
that  the SED  is only  reproducible  with a  young synthetic  stellar
population and it is incompatible with an evolved population.

The  starburst scenario  is also  supported  by the  results obtained  from
fitting   the   K96  observational   templates.    As   it   is  shown   in
Table~\ref{empfitresults}, by  far the smallest values  of $\chi^2/dof$ are
obtained with blue, low extinction,  starburst templates (Stb2 and a Stb1). 
The rest  of the  empirical templates, especially  the early  type galaxies
(S0,  E and  B), show  a high  degree of  discrepancy with  the photometric
points ($\chi^2/dof > 23.71$, see Table~\ref{empfitresults}).

Therefore, combining  the morphological  information and the  synthetic and
empirical SED fits we conclude that  the GRB~000418 host galaxy SED is best
reproduced  by a  Stb template.   This agrees  with the  independent result
reported  by  Bloom et  al.   (\cite{Bloo03})  who  based on  the  relative
intensities   of  the   [\ion{O}{II}],  \ion{He}{I},   [\ion{Ne}{III}]  and
\ion{H}{$\gamma$}  lines, conclude  that  the host  is  a starburst  galaxy
rather than a LINER or a Seyfert 2 galaxy.

The  lack of  underlying peripheric  bright  knots of  star formation  (see
Sect.~\ref{ellipse}) supports a model  with one dominant nuclear starburst. 
This  fact makes our  photometric points  a suitable  input for  Hyperz, as
multiple contemporaneous episodes of star  formation can not be fitted with
this code.

\subsubsection{Metallicity}

In  the upper sub-table  of Table~\ref{synfitresults}  we show  the results
obtained when  solar metallicity  is assumed for  the host galaxy.   In the
middle  sub-table the  results are  displayed when  the metallicity  is not
fixed.  In  such case  the stars eject  heavy elements to  the environment,
enriching the ISM where new generation of stars are continuously born.  The
effect of  the ISM  enrichment is  expected to be  maximum for  Im galaxies
(where the  SFR is constant)  and negligible for  instantaneous starbursts,
where  all  the  stars  are  modeled   to  be  formed  at  the  same  epoch
(instantaneous SFR idealised by a delta function).

As it is shown the results  of both sub-tables are basically the same.
Even if the  metallicity is left as a free  parameter the Stb template
is the  one providing the  most satisfactory fits.  Thus,  we conclude
that the metallicity of the host is consistent with Solar metallicity,
but that the metallicity is not strongly constrained by our analysis.

\subsubsection{Extinction}
\label{ext}

According to the empirical SED templates, the blue SED of the host can only
be  roughly reproduced  with low  extinction starburst  galaxies (templates
Stb1 and Stb2).  As it can be seen in  Table~\ref{empfitresults} there is a
clear correlation  between the goodness  of the fit  and the colour  of the
template: the  bluer the colour  the lower $\chi^2/dof$.  To  translate the
value of  $E(B-V)$ to  $A^{global}_{\rm V}$ we  adopt the value  of $R_{\rm
  V}=A^{global}_{\rm V}/E(B-V)=4.05\pm0.80$ proposed for starburst galaxies
(Calzetti et  al.  \cite{Calz00}).   The best fit  is achieved with  a Stb2
template, which  shows by definition a  fixed colour excess  $E(B-V) = 0.15
\pm 0.05$ ($A^{global}_{\rm V}=0.61 \pm 0.24$ following Cal00).

The  host  galaxy extinction  $A^{global}_{\rm  V}$  derived from  the
synthetic    SEDs    range    from    0.12   to    1.47    mag    (see
Table~\ref{synfitresults}).   The derived $A^{global}_{\rm  V}$ values
are  mostly  dependent  on  the  assumed extinction  law,  and  almost
independent  of the  IMF and  metallicity.  The  Cal00  extinction law
predicts the  existence of a very  blue, young (age $\sim$  4 Myr) and
extincted   stellar  population   ($A^{global}_{\rm  V}   \sim  1.4$).
However, the  inferred high $A^{global}_{\rm  V}$ is not  supported by
the empirical template fits, so this solution seems quite unlikely and
for the further  discussion it will only be  considered as a secondary
scenario.

For  illustrative purposes,  we have  included  in the  lower sub-table  of
Table~\ref{synfitresults}   the    results   obtained   when    we   impose
$A^{global}_{\rm V}=0$.  As  it is shown these fits  are still satisfactory
($\chi^{2}/dof \sim 1.6$) and the photometric redshift is very close to the
spectroscopic  one ($z  \sim 1.02$),  implying  that the  assumption of  no
extinction is not in strong conflict with the data.

In conclusion, the synthetic and empirical templates yield similar results;
the  fits  to  the  GRB~000418  host  galaxy SED  lead  to  a  low/moderate
$A^{global}_{\rm V}$ ranging from $0.12$ (e.g. Sc86-Sea79 synthetic SED) to
$0.61$  mag (Stb2  empirical template).   In the  following we  will assume
$A^{global}_{\rm  V}=0.36 \pm  0.25$.  This  value is  consistent  with the
extinction along the line-of-sight to the OA ($A^{local}_{\rm V} = 0.65 \pm
0.25$).  Given  that $A^{local}_{\rm V}$  and $A^{global}_{\rm V}$  are not
very  different we  conclude that  the GRB  apparently did  not occur  in a
region with more than average extinction.

\subsubsection{Star Formation Rate}
\label{sfr}

The  UV  continuum  emission  is  dominated  by  bright,  short-lived,
main-sequence O  and B stars.  According  to Kennicutt (\cite{Kenn98})
the  rest-frame  2800  \AA~  flux  is  directly  proportional  to  the
star-formation rate  in the  host galaxy.  Therefore,  the SFR  can be
estimated from this part of the spectrum.

When this  diagnostic method  is applied to  starburst galaxies  it is
affected by two  main uncertainties. The first one  is the sensitivity
of the  estimator to extinction, which  has a strong impact  on the UV
region. The  young massive stars can  be located in  regions where the
$A^{local}_{\rm V}$  properties differ substantially  from the average
host  extinction  (derived from  the  SED  fitting),  so a  proper  UV
correction is difficult to  obtain.  The second uncertainty comes from
the  fact  that  the  expression given  by  Kennicutt  (\cite{Kenn98})
relating  the SFR$_{UV}$  and  the UV  luminosity  ($L_{UV}$) is  only
strictly valid  for galaxies with continuous star  formation over time
scales of  $10^8$ years or  longer.  The SFR$_{UV}$/$L_{UV}$  ratio is
significantly  lower   for  younger  populations   such  as  starburst
galaxies.   Hence, these  two systematic  uncertainties  have opposite
effects on the derived SFR.

At $z=1.118$ the  rest-frame 2800 \AA~ corresponds to  5930 \AA, so it
is bracketed between our $V$ and $R$ bands.  The flux at 5930 \AA~ has
been  determined  by fitting  a  power law  spectrum  to  the $V$  and
$R$-band photometric  points (once corrected  for Galactic extinction)
and then interpolating it at 5930 \AA.  Finally, the interpolated flux
has been dereddened according  to the corresponding extinction law and
the $A^{global}_{\rm V}$ value inferred with the synthetic SED fitting
(see column  7 of Table~\ref{synfitresults}).  The  resulting SFRs can
be seen in Table~\ref{synfitresults}.

 As   it   is   shown,   the   $SFR_{UV}$   estimates   displayed   in
Table~\ref{synfitresults} (once  the unlikely SED  solutions have been
discarded)  are $\sim  7$ times  lower than  the $SFR_{[\ion{O}{II}]}$
estimates  based  on  the  [\ion{O}{II}] line  flux,  $L[\ion{O}{II}]$
(Bloom et al.  \cite{Bloo03}).   The main reason for this disagreement
is     the     high     $SFR_{[\ion{O}{II}]}/L[\ion{O}{II}]$     ratio
($5\times10^{-41}$, given by Kennicutt \cite{Kenn92}) assumed by Bloom
et     al.      (\cite{Bloo03}).       If     the     more     updated
$SFR_{[\ion{O}{II}]}/L[\ion{O}{II}]$    ratio    ($1.4\times10^{-41}$)
reported by Kennicutt (\cite{Kenn98};  the reference used in our study
to estimate $SFR_{UV}$) is applied, then the $L[\ion{O}{II}]$ measured
by Bloom et al.  (\cite{Bloo03}) corresponds to $SFR_{[\ion{O}{II}]} =
15.4 M_{\odot}$  yr$^{-1}$ (with a  systematic error of  $\sim 30\%$),
close to our $SFR_{UV}$ estimate.

A  second order  parameter that could  explain the factor  of two
still present between both $SFR$ estimates might be the naive scenario
we assumed for the local  extinction correction.  In principle, if the
local   extinction  estimation   was   perfect,  the   $UV$  and   the
$[\ion{O}{II}]$   diagnostic  methods   should  yield   same  results.
However, if (part or most  of) the massive star population responsible
of  the  UV radiation  is  embedded  in  extincted regions,  then  the
$A^{global}_{\rm V} = A^{local}_{\rm V}$ approximation (used to derive
the  $SFR_{UV}$ values  displayed in  Table~\ref{synfitresults}) would
underestimate the $UV$ flux.

Additionally, the  reddening correction for $SFR_{[\ion{O}{II}]}$
has  to  be  carried  out  at  $H_{\alpha}$  (6563  \AA)  and  not  at
[\ion{O}{II}] (3727  \AA), given  the manner the  [\ion{O}{II}] fluxes
were calibrated (Kennicutt  \cite{Kenn92}, \cite{Kenn98}).  This makes
the extinction correction for the $SFR_{UV}/SFR_{[\ion{O}{II}]}$ ratio
even higher.

Considering  the   above  mentioned  uncertainties   and  limitations,
intrinsic to  both $SFR$ diagnostic  techniques, we conclude  that the
$SFR_{UV}$ based  on the  our $UBVRIZJ_sK_sCL$-band SED  is compatible
with   the   $SFR_{[\ion{O}{II}]}$   estimated   by   Bloom   et   al.
(\cite{Bloo03}).

\subsubsection{Stellar population age}

Information on  the stellar  population age can  be inferred  from the
strength of the  4000~\AA~ jump by means of  synthetic templates.  The
age of the derived Stb synthetic SEDs (the rest of synthetic templates
have  been considered  much  more unlikely;  see Sect.~\ref{type}  and
\ref{ext}) range from 0.053 to 0.064  Gyr. An age of $0.059 \pm 0.006$
Gyr  (the mean  of value  of 0.053  and 0.064)  corresponds (basically
independent of the metallicity) to the  lifetime of a star with $7 \pm
1 M_{\odot}$ (see Table 14 of Portinari et al.  \cite{Port98}).

\subsubsection{Photometric redshift}

Given that the spectroscopic redshift is known, the photometric redshift is
only  used  to  check the  internal  consistency  of  the fitted  SEDs.   A
systematic  comparison  performed  between  spectroscopic  and  photometric
redshifts inferred with Hyperz for a sample of 10 GRB hosts shows a typical
redshift dispersion  $\sim$0.1 (Christensen et  al.  \cite{Chri02}).  Thus,
we  conclude that  the synthetic  SED  fits reproduce  reasonably well  the
spectroscopic  redshift   of  our  host   galaxy  (see  fourth   column  of
Table~\ref{synfitresults}).

  We  have checked  the  possible impact  that  the poorly  determined
  U-band magnitude (error 0.3 mag)  might have in the determination of
  the  photometric redshift.   Thus,  we have  repeated  all the  fits
  displayed  in Table~\ref{synfitresults}  excluding  the U-band  host
  magnitude.  The derived photometric redshifts differ less than $2\%$
  (achieved for the SOL-MiSc79-Fit86  subfamily of templates) from the
  ones obtained with the entire $UBVRIZJ_sK_sCL$-band SED.  This small
  variation is the result of the weighted calculation of $\chi^2/dof$,
  which weights each band according to the square of its corresponding
  photometric  error inverses  (see Hyperz  manual, Bolzonella  et al.
  \cite{Bolz02}).   In  the  same  manner  the impact  of  the  U-band
  magnitude  on   rest  of   the  inferred  variables   (galaxy  type,
  metallicity, age, template, $A^{global}_{\rm V}$) is also negligible
  for  the further  discussion (the  maximum impact  corresponds  to a
  variation of one $A^{global}_{\rm V}$ grid step, 0.03 mag).

The  reliability of our  empirical templates  fits have  been also  tested. 
Leaving the redshift as a free parameter, (and filtering the spurious local
$\chi^2/dof$  minimum frequently found  at $z=0$),  only the  Stb2 template
yields a reasonable photometric  redshift ($z=1.272$, consistent within the
expected  redshift  dispersion  of  $\Delta  z \sim  0.1$).   The  rest  of
templates, specially  the early and mid types,  give redshifts inconsistent
with the  spectroscopic one.   This fact supports  that the  Stb2 empirical
template (see Sect.~\ref{ext}) is the optimum one to reproduce our data.

\subsubsection{Luminosity of the host galaxy}

The  absolute  $B$-band  magnitude of  the  host  at  $z=1.118$ is  $M_B  =
-20.6\pm0.1$.   Lilly  et  al.   (\cite{Lill95})  show  that  $M^{\star}_B$
depends  on the colour  and the  redshift of  the galaxy.   This luminosity
evolution is specially  relevant for blue galaxies at $z  \sim 1$ (like the
GRB~000418  host  galaxy),  where  $M^{\star}_B$ ranges  from  $-21.22$  to
$-22.93$   (rescaling  the   $M^{\star}_B$   values  of   Lilly   et  al.   
(\cite{Lill95}) to our cosmology). Although the $M^{\star}_B$ value of blue
galaxies  is very  uncertain, the  trivariate luminosity  function  (LF) of
Lilly et al.   (\cite{Lill95}) suggests that the value  of $M^{\star}_B$ of
blue galaxies is $ < -20.6$. Therefore, we conclude that the host is likely
a subluminous galaxy.

\section{Discussion: fitting all the pieces of the puzzle}

Several characteristics of the GRB~000418 host galaxy are difficult to
reconcile, in  particular: {\em i)}  a high reddening is  expected for
sub-mm luminous  galaxies (see Le  Floc'h et al.  2003).   However, we
find  that  GRB~000418   occurred  in  a  blue  host   galaxy  with  a
low/moderate   extinction;    {\em   ii)}   A$^{local}_{\rm    V}$   /
A$^{global}_{\rm V} \sim 1$; {\em iii)} SFR$_{mm}$/SFR$_{UV} \sim 50$.
Below we discuss several scenarios  that could help to reconcile these
observations.

\subsection{Obscured stellar population}
\label{hidden}
The  Sub-mm/radio emission could  trace an  obscured population  of massive
stars that could be undetectable  at optical/NIR wavelengths. This has been
suggested in the case of the  host galaxy of the dark GRB~000210 (Gorosabel
et al.  \cite{Goro03}).

Given  that SFR$_{UV}$/SFR$_{mm}$  $\sim  0.02$, the  probability that  the
progenitor belongs to the obscured population is $\sim 98\%$ (assuming that
the probability  of making a  GRB is only  proportional to the SFR  and not
other parameters as  e.g. the metallicity).  However, opposite  to the case
of GRB~000210,  GRB~000418 was  not dark, so  its progenitor either  had to
belong to the remaining $\sim 2\%$ unobscured stellar population or the GRB
destroyed the dust along the line of sight.

\subsection{Nuclear radio/sub-mm supernovae or AGN  activity}
\label{agn}
Another  possibility is that  the radio/sub-mm  emission comes  from a
high nuclear  radio supernovae  rate ($>1$ SN  yr$^{-1}$) or  from the
activity associated  with an AGN.   Thus, the radio/mm  emission would
not invoke an optically  hidden stellar population and the discrepancy
between SFR$_{UV}$ and SFR$_{mm}$ would be naturally solved.

An  appreciable amount  of  starbursts ($\sim  40\%$) contain  compact
radio cores (Kewley et al.  \cite{Kewl99}).  These compact radio cores
may be  originated by obscured AGN  or by complexes  of luminous radio
supernovae   from  an   active   nuclear  starburst   (Smith  et   al.
\cite{Smit98a}).  The  GRB~000418 host galaxy would  resemble the case
of Arp 220, an active star forming galaxy (SFR $\sim 50-100~M_{\odot}$
yr$^{-1}$)   which  shows  a   compact  radio   core  (Smith   et  al.
\cite{Smit98b}).  The  additional radio  source detected by  Berger et
al. (\cite{Berg03})  could also be  related to AGN activity  (e.g. the
hot spot of a radio jet).

\section{Conclusions}
\label{conclusions}

The  analysis  of optical/NIR  observations  presented  in this  paper
confirms that the GRB~000418 host  is a starburst galaxy.  This result
has  been independently  achieved by  fitting synthetic  and empirical
templates  to  the  photometric   points.   This  conclusion  is  also
consistent  with  the   morphological  information  derived  from  the
HST/STIS images, where the host is  seen as a blue compact galaxy with
no  evidence for  more widespread  star formation.   The  more natural
scenario would be a nuclear  starburst that harbour a young population
of stars  where the  GRB was originated.   The reported offset  of the
afterglow  respect  to   the  galaxy  nucleus  ($0.023\pm0.064^{\prime
\prime}$;  or   a  projected  distance  of   $0.202\pm0.564$  kpc)  is
consistent with this hypothesis (Bloom et al.  \cite{Bloo02}).

The synthetic SED fits are consistent with a young stellar population.
The predicted  host galaxy extinction, stellar age  and star formation
rate  depend  on  the  assumed  extinction  law.   Two  synthetic  SED
solutions are consistent with our photometric points: i) age = $59 \pm
6$ Myr, $A^{global}_{\rm  V} = 0.15 \pm 0.03$,  SFR$_{UV} \sim  8
M_{\odot}$ yr$^{-1}$, progenitor mass = $ \sim 7 M_{\odot}$; ii) age =
$4 \pm 1$ Myr, $A^{global}_{\rm V} = 1.41 \pm 0.06$, SFR$_{UV} \sim 55
M_{\odot}$  yr$^{-1}$, progenitor  mass  = $\sim  60 M_{\odot}$.   The
second  solution is  not  likely  since it  is  inconsistent with  the
independent results obtained with  the empirical templates, however it
can not be completely discarded.  In any case, both solutions are much
easier   to   accommodate    in   the   collapsar   context   (Woosley
\cite{Woos93a}; Paczy\'nski \cite{Pacz98})  than in the binary merging
scenario (Eichler et al.  \cite{Eich89}).

The global  extinction of the host  (defined as the  averaged value of
the  ones obtained  with  the synthetic  and  empirical templates)  is
$\sim0.4$ mag, similar to the one  measured along the line of sight to
the afterglow.  It  is consistent with a rather  homogeneous ISM, with
no large  density fluctuations, at least around  the progenitor.  This
hypothesis  would be  supported  by the  smooth  and symmetric  radial
profile  inferred  from the  HST  data.  So  the  GRB  might have  not
occurred in a extremely overdensed region of the host.

The inferred  SFR$_{UV}$ ($\sim 8 M_{\odot}$ yr$^{-1}$)  is two orders
of   magnitude  lower   than  the   one  inferred   from  sub-mm/radio
measurements.

Several ISM configurations have been proposed to explain this disagreement.
In most  of them an obscured  population of massive stars  (only visible in
the  sub-mm/radio range)  is invoked.   An alternative  way to  explain the
radio emission could be the presence  of a nuclear radio core originated by
an AGN or by complexes of extremely luminous radio supernovae.

\section*{Acknowledgments}
J.  Gorosabel acknowledges  J.-M. Miralles, M.  Bolzonella  and R.  Pell\'o
for helpful assistance with  Hyperz.  J.M.  Castro Cer\'on acknowledges the
receipt of a FPI doctoral  fellowship from Spain's  Ministerio de Ciencia y
Tecnolog\'\i  a.  J.P.U.   Fynbo  acknowledges support  from the  Carlsberg
Foundation.  STH acknowledges support from  NASA LTSA grant NAG5-9364. Part
of the  observations presented in  this  paper were obtained  under the ESO
Programmes 165.H-0464(I),  265.D-5742(C)  (granted to  the  GRACE team) and
67.B-0611(A) (public  data retrieved from ESO archive).    Part of the data
presented in this  paper were acquired with ALFOSC,  which is owned by  the
Instituto de Astrof\'{\i}sica de  Andaluc\'{\i}a (IAA) and operated  at the
Nordic  Optical  Telescope under agreement between   IAA and NBIfAFG of the
Astronomical Observatory of Copenhagen.  We thank our anonymous referee
by fruitful and constructive comments.

\end{document}